\documentclass[prl,aps,twocolumn,groupedaddress,showpacs,floatfix]{revtex4}

\usepackage{amsmath,amssymb,multirow,epsfig,bm}

\newcommand{\etal}{{\it et al.,\;}}
\newcommand{\beq}{\begin{equation}}
\newcommand{\eeq}{\end{equation}}
\newcommand{\bea}{\begin{eqnarray}}
\newcommand{\eea}{\end{eqnarray}}

\newcommand{\benn}{\begin{displaymath}}
\newcommand{\eenn}{\end{displaymath}}

\begin{document}

\title{\bf Molecular transitions in Fermi condensates
 }

\author{ Aurel Bulgac$^1$ and George F. Bertsch$^{1,2}$ }
\affiliation{$^1$ Department of Physics, University of
Washington, Seattle, WA 98195--1560, USA}
\affiliation{$^2$Institute of Nuclear Theory,
University of Washington, Seattle, WA 98195-1550, USA}

\date{\today}

\begin{abstract}

We discuss the transition of fermion systems to a condensate of Bose
dimers, when the interaction is varied by use of a Feshbach
resonance. We argue that there is an intermediate phase between the
superfluid Fermi gas and the Bose condensate of molecules, consisting
of extended dimers.

\end{abstract}



\maketitle


The remarkable experimental advances in atomic trap physics now
permits one to study Fermi gases of atoms over a broad range of
couplings from weak to strong.  In particular, two-component Fermi
systems have been made and studied using the two alkali metal isotopes
$^{40}$K \cite{cindy,regal,collisions} and $^6$Li
\cite{others,selim,clouds}.  In both cases, the Feshbach resonance
phenomenon has been exploited to vary the coupling strength of the
interaction between atoms and study the consequent many-particle
properties.  In both cases there is a closed channel near threshold
who energy can be tuned with respect to the open two-body channels.  A
particularly interesting feature is the formation of molecules or
dimers as the effective interaction strength is increased.  There are
many questions than can be posed about a coexisting phase of bound
pairs and that we would like a theory to address.

Here we want to argue that the theoretical situation is rendered more
complex by the fact that there are fact two kinds of pair states that
must be treated in any theory that applies to the full range of
couplings. These two kinds of pairs are spatially different and have
different quantum numbers as well.  The first bound state appears when
the inverse scattering length passes through zero (via the tuning of
the Feshbach resonance).  We shall call it the {\it halo dimer}, in
analogy with somewhat similar {\it halo nuclei} \cite{halo}. A halo
dimer is spatially extended and has only a small overlap with the
closed-channel Feshbach resonance.  But as the closed-channel
resonance is tuned to lower energies, its energy eventually becomes
negative and can be identified with the pair state.  We present below
a simple model that shows how this two-step transition from the
continuum pair states to the molecular bound state takes place,
deriving the relevant size and energy scales.  We then discuss the
experimental signatures of the halo dimer state.  Finally, we assess
the theoretical approaches that have been applied to treat the
many-particle system in the halo dimer domain.  There are a number
calculations in the literature that ignore the special character of
these state, and we believe that such models lead to a number of 
falsifiable predictions.

We consider here a simple model of the two-body physics use a
two-channel atomic Hamiltonian of the form
\beq
 \left [
\begin{array}{cc}
H_{11}(r)& V_{12}(r)\\   V_{21}(r)&  H_{22}(r)
\end{array}
\right ]
\left [ \begin{array}{l}  u(r)\\ v(r) \end{array} \right ]
= k^2
\left [ \begin{array}{l}  u(r)\\ v(r) \end{array} \right ],
\eeq
where $r$ is the atom-atom separation. We shall use the units
$\hbar=m=1$ and here $u(r)$ describes the open two-atom channel and
$v(r)$ the closed (molecular) channel. It is implied that while
$V_{11}(r) \rightarrow 0$ and $V_{12}(r)=V_{21}(r)\rightarrow 0$ when
$r \rightarrow \infty$, the closed channel potential tends to a
positive constant $V_{22}(r)\rightarrow U_0>0$. Since we are
interested in very low energies, only the $s$-wave is considered here
and we show only the radial parts of the wave functions. If one were
to apply this model to a system of $^6$Li atoms, in the open channel
at magnetic fields near the Feshbach resonance the two valence
electrons of the two $^6$Li atoms would be anti-parallel to the
magnetic field (triplet state), while in the closed channel the two
electron spins would be anti-parallel to each other (singlet state).
As in all previous references
\cite{falco,ohashi,holland,griffin,eddie} we shall assume that in the
closed channel there is only one state close to the two-atom
threshold. If the wave function of this state is $\phi_0(r)$ with an
energy $\kappa_0^2$ with respect to the two-atom threshold, one can
easily show that the two-channel problem simplifies somewhat and the
wave functions read
\bea
& &\!\!\! \!\!\!
 u(r) = u_0(r) +
\left \langle r\left | \frac{1}{k^2-H_{11}}V_{12}\right | \phi_0\right \rangle
\frac{\langle\phi_0|V_{21}|u\rangle}{k^2-\kappa_0^2},\\
& & v(r) =
\phi_0(r)\frac{\langle\phi_0|V_{21}|u\rangle}{k^2-\kappa_0^2},\\
& & H_{11}(r) u_0(r) = k^2u_0(r),\\
& & H_{22}(r)\phi_0(r) = \kappa_0^2 \phi_0(r),
\eea
For the sake of simplicity, we shall assume that in the open channel
the atoms are free (the same assumption was made in
Ref. \cite{falco}), that $\phi_0(r) = \sin (\pi r/r_0) \sqrt{2/r_0}$
for $0\le r \le r_0$ and zero otherwise and that
$V_{12}(r)=V_{21}(r)=g\delta(r-r_1)$, with $r_1=r_0/2$. Physically,
the parameter $r_0$ is of the order of the van der Waals length $
r_0\approx (C_6 m/\hbar^2)^{1/4}$ and we shall consider only such
energies for which $kr_0\ll 1$. After some simple manipulations one
arrives at the following form of the two-atom wave function in the
open channel (using $u_0(r) = \sin(kr)$)
\beq
u(r) = \sin (kr) +
\frac{g^2\exp (ikr_>)\sin (kr_<)}{\kappa_0^2-g^2 - ig^2 kr_1},
\label{eq:u}
\eeq
where $r_>=\max(r,r_1)$ and $r_<=\min(r,r_1)$ and  terms of order
${\cal{O}}(k^2)$ have been neglected. One can then easily show  that
\bea
& & -\frac{1}{a} = \frac{\kappa_0^2-g^2}{g^2r_1}, \label{eq:a}\\
& & u(r) = \exp(i\delta )\sin k(r-a), \quad {\mathrm{if}} \quad r\ge
r_1,
\eea
where $\tan \delta = -ka$.  In particular, exactly at the resonance
$u(r) = i \cos(kr)$ for $r\ge r_1$. Far away from the Feshbach
resonance, the open channel wave function would be approximately equal
to $u(r)=\sin (kr)$ instead (for $r\ge r_0$).  By the nature of the
problem at hand we have $\kappa_0r_0={\cal{O}}(1)$ and
$gr_0={\cal{O}}(1)$.  Only if $\kappa_0$ and $g$ are comparable in
magnitude one can attain the regime when $a=\pm \infty$.  For the
experiments reported, the relative energy of the Feshbach resonance
and thus the magnitude of $\kappa_0^2$ is controlled by a magnetic
field.  With fine tuning one can make the numerator in
Eq. (\ref{eq:a}) very small, and in this way attain the regime $|a|\gg
r_0$, even though all the parameters in the equation above are of
order $1/r_0$ or $r_0$ respectively. As an order of magnitude estimate
for these constants one can use
${\cal{O}}(\kappa_0^2)={\cal{O}}(g^2)=2\mu_B B_0 m/\hbar^2$, where $m$
is the atomic mass, $\mu_B$ is Bohr magneton and $B_0$ is the value of
the magnetic field where $1/a=0$. For both $^6$Li and $^{40}$K one
readily obtains that ${\cal{O}}(\kappa_0r_0) ={\cal{O}}(g r_0) = 1$.

Because of the coupling between the two channels, there is now a pole
of the scattering amplitude at $k_0=i/a$, as may be seen from
Eq. (\ref{eq:u}).  In different terms, by means of a magnetic field
one controls the logarithmic derivative of the open channel wave
function near the origin, more exactly near $r=r_0$. In a finite
density medium it might naively appear that an infinite value of the
scattering length would be meaningless. However, that should be
interpreted rather as the logarithmic derivative of the open channel
wave function at $r=r_0$, namely $d \ln u(r)/dr|_{r=r_0}=-1/a$, or in
more physical terms, as the relative momentum with which the two atoms
emerge after interacting at short distances. When the two-atom system
is far from the Feshbach resonance the typical relative momentum with
which the two atoms emerge after interacting at distances smaller than
$r_0$ is $\hbar/r_0$. The special situation, which is achieved by
bringing the two atoms exactly at the Feshbach resonance, is to insure
that they emerge from the interaction region with an essentially
vanishing relative momentum. In a certain sense that amounts to an
ultimate further cooling of the relative atomic motion to its minimum
and that is what makes the physics of atoms under these conditions
particularly exciting. In a sense, two-atom collisions at short
distances do not bring in any momentum.

Let us put this two-atom system in a spherical cavity of radius $R\gg
r_0$. In principle one would have to specify the boundary condition
for $u(r)$ at $r=R$, which would lead to energy quantization. The
specific nature of this energy quantization (e.g.  Neumann {\it vs}
Dirichlet boundary conditions) is qualitatively unimportant.  By
choosing the radius of this spherical cavity one can simulate a Fermi
gas of various densities, with $k_F\propto 1/R$, and in particular
choose the regime with $k_F|a|\gg 1$, when according to the authors of
Ref. \cite{falco} the two atoms should be with probability essentially
one in the small size molecular state. One has simply to estimate the
ratio of the probability to find the atoms in one or another channel,
and one easily finds that
\bea
\!\!\! \!\!\!
\int dr |v(r)|^2& = &\frac{k^2a^2}{(k^2a^2+1)g^2r_1}\leq
\frac{1}{g^2r_1}
={\cal{O}}(r_0), \label{eq:v2}\\
\!\!\! \!\!\!
\int_0^R dr  |u(r)|^2 &\approx &  \int_{r_0}^R dr \sin^2 (kr +\delta)
={\cal{O}}(R). \label{eq:u2}
\eea
Consequently, the relative probability to find the two atoms in the
molecular state is of the order ${\cal{O}}(r_0/R)
={\cal{O}}(k_Fr_0)\ll 1$ for a dilute Fermi gas, $nr_0^3\ll 1$.  

One can continue the
argument in the region of negative detuning, when $a>0$, and easily
convince oneself that as long as $a\gg r_0$, the relative probability to
find the two atoms in the closed channel is small. That was discussed
in Ref. \cite{braaten} and shown in another explicit calculation in
Ref. \cite{kohler}. Only when the scattering amplitude becomes of the
order of the radius of the interaction (van der Waals length) the
probability to find two atoms in the closed channel becomes comparable
with the probability to find the same atoms in the open channel.
The various regimes of the coupling strength are shown in Table I.

\begin{table}
\caption{ Character of the condensate as a function of the inverse
scattering length $a^{-1}$ in various in intervals, 
the approximate boundaries of these intervals being shown in the second row.
The total electron spin and spin projection $(S,S_Z)$ along the
magnetic field for various pairs are shown in the last row. }
\begin{tabular}{|c|c|c|c|c|}
\colrule
\multicolumn{3}{|c|} {$a^{-1}>0$} & \multicolumn{2}{|c|} {$ a^{-1}<0$} \\
\colrule
\multicolumn{3}{|c|} {$+\infty$ \hfill $r_0^{-1}$ \hfill $k_F$ \hfill 0} & 
\multicolumn{2} {|c|}{0 \hfill $k_F$ \hfill $-\infty$} \\
\colrule
\qquad\qquad\quad  &  halo       & \qquad\qquad\quad  & BCS      &  BCS  \\
molecules  & \quad  dimers \quad & {\large ?}         & strong   & weak \\
           & (+ atoms ?\cite{bbf})&          & coupling & coupling\\
\colrule
(0,0)      &  (1,-1)             & (1,-1)             &  (1,-1)  & (1,-1) \\
\colrule
\end{tabular}
\end{table}

Next we discuss experimental observables to distinguish between the
two kinds of bound states.  {\it i)} One signature is magnetic.  The
experiments are typically carried out in magnetic fields large
compared to the hyperfine splitting.  The continuum states are thus
polarized with respect to the electron spin, with the nuclear spin
mainly responsible for the second component of a two-component Fermi
gas.  Thus the electron spin wave function of the halo dimer is
largely S=1.  The closed channel, on the other hand, is typically well
described with a spin-singlet electron wave function.  In the
experiment of Ref. \cite{selim}, the spin of the bound pair was
measured (see the insert of Fig. 2), and it was indeed found that its
value was large down to values of the magnetic field well below the
value at which the scattering length changes sign. This result is in
perfect agreement with microscopic understanding of the Feshbach
resonance in $^6$Li \cite{theory}. {\it ii)} Another signal of the
character of the pair state in the condensate is in the particle loss
rate of the system.  When the condensate is prepared with the atoms in
the lowest hyperfine states, the only inelastic processes are
three-body collisions in which one of the pairs goes into a deeply
bound molecule.  Such processes are more probable when three particles
are simultaneously in each other's range of interaction.  The
observations of Ref. \cite{selim}, see Fig. 4, confirmed by a separate
experiment Ref. \cite{collisions}, in fact show that the loss rate
becomes large under the same conditions that the bound pair develops a
singlet character.  This observation also agrees with the theoretical
expectation of Ref. \cite{petrov}.  {\it iii)} We mention one more
piece of evidence, this one more indirect.  As several experimental
groups have demonstrated \cite{clouds} when the scattering length is
positive (thus on the BEC side of the Feshbach resonance) and when
$na^3 < 1$ the sizes of the atomic clouds agree very well with theory
based on halo dimers, which predicts that the interaction between them
can be characterized by a scattering length of magnitude $0.6 a$
\cite{bbf,petrov}.  If the pairs were in closed-channel molecular
state, their scattering length would have the order of magnitude of
their size ${\cal{O}}(r_0)$ independent of the atom-atom scattering
length $a$.

We now turn to theoretical models for the condensation of trapped
fermions into molecules.  There have been a number of recent
calculations \cite{falco,ohashi,holland,griffin}, based on the model
of Timmermans \etal \cite{eddie} that assume a direct transition
between the fermions and a molecular bound state.  In Ref.
\cite{falco} it was claimed that a dilute atomic Fermi gas near a
Feshbach resonance undergoes a crossover into diatomic molecules of
relatively small sizes on the BCS side of the resonance, when the
atomic scattering length is still negative and the molecular energy
level lies in the two-atom continuum. The authors of Ref. \cite{falco}
describe the results of the recent experiment of Regal {\it et al.}
\cite{regal}, by the way of this conversion of atoms into diatomic
molecules.  Another recent preprint makes a similar claim, that
exactly this process occurs on the BCS side of the Feshbach resonance
\cite{ohashi}, although this particular work deals with different
properties of such systems.

The statement we take issue with here was formulated extremely
succinctly by the authors of Ref. \cite{falco}. One considers a
uniform dilute Fermi gas of number density $n=k_F^3/3\pi^2$, near a
Feshbach resonance at positive detuning ($a<0$), when the energy of
the molecular state is $\epsilon_m\simeq \hbar^2/ma^2>0$. With respect
to the two-atom continuum this molecular state is unbound.  Then if
$k_F|a|>1$ a fraction of these atoms converts into molecules with a
number density, see Ref. \cite{falco} and also
Refs. \cite{ohashi,griffin} where a similar result is quoted,
\beq
n_{mc}\simeq \frac{n}{2}\left [ 1 - \frac{1}{(k_F|a|)^3}\right ].
\label{eq:nmc}
\eeq
Naively this results seems obvious, since as soon as the Fermi energy
exceeds the (positive) molecular state energy, the system can only
lower its total energy by converting into molecules. These molecules,
being bosons in character, can all have zero center of mass kinetic
energy, and thus the Fermi energy is lowered until it becomes equal to
$\epsilon_m\simeq \hbar^2/ma^2$ and further conversion of atoms into
molecules becomes energetically impossible. Applying this to the
two-body problem, the transition would take place discontinuously as
the inverse scattering length approaches zero from the BCS side.  But
as we saw previously, the formation of halo dimers is a smooth process
with the two-particle wave function changing smoothly as the inverse
scattering length goes through zero. It is also abundantly clear that
such models \cite{falco,ohashi,holland,griffin} would predict the
wrong magnitude for the magnetic moments of such Fermi clouds near the
Feshbach resonance. If indeed the BEC of small size molecules would
occur in this regime, the prediction would be that the magnetic moment
of such pairs would be vanishing, while experiment clearly shows that
not to be the case \cite{selim}.

There is still one possibility, when the mechanism suggested in Refs.
\cite{falco,ohashi,holland,griffin} could prove indeed correct and
near a Feshbach resonance such a system would condense into molecules
of relatively small size ${\cal{O}}(r_0)$. This situation would occur
if the coupling constant $g$ becomes unnaturally small. The relative
probability that two atoms spend most of the time in a small size
configuration, according to Eqs. (\ref{eq:v2} - \ref{eq:u2}) would be
$\propto 1/(g^2r_1R)\approx k_Fr_0 /(gr_0)^2$, which could become
accidentally large, even though the system is still dilute, in the
sense that $k_Fr_0\ll 1$. This would happen if the Feshbach resonance
would be accidentally extremely close to the atom-atom threshold.
Whether this mechanism can be indeed realized in experiments with
dilute atomic Fermi clouds, is a question that remains to be
investigated.

Since the halo dimers occupy the same sector of the two-particle
Hilbert space as the plane waves of the Fermi gas, it is reasonable to
ask whether a theory can be found that does not introduce a closed
channel resonance explicitly.  In fact, the BCS theory has the degrees
of freedom to describe the many-particle system over the full range of
couplings we consider here \cite{exact}.  One may show \cite{leggett}
that the BCS theory contains the correct description of the gas of
halo dimers in the limit that $a \ll 1/k_F$ and positive.  Namely, the
BCS chemical potential $\mu$ is related to the dimer energy $E_d$ by
\beq
\mu = \frac{E_d}{2}=-\frac{\hbar^2}{2ma^2}.
\eeq
On the other hand, the BCS theory is quantitatively incorrect on at
least two properties.  The pairing gap $\Delta$ is reduced by
polarization effects from the BCS value by about a factor of $\approx
2$ both in the weak \cite{gorkov} and strong coupling limit
\cite{carlson}.  Also, the dimer-dimer scattering amplitude is
predicted to be $2a$ in the BCS theory \cite{randeria}, while a more
detailed treatment gives $0.6 a$ \cite{bbf,petrov}. Still, it is still
possible to use the a BCS framework for an effective theory,
renormalizing the couplings to reproduce these interactions, in the
framework of the Superfluid LDA (SLDA) \cite{abyy}.

We would like to thank H.-W. Hammer and E. Timmermans and especially
B. DeMarco for a series of discussions, which help better appreciate
some of the prevailing attitudes in the community and H.-W. Hammer for
reading and making a series of suggestions on an earlier version of
the manuscript.  This work was supported in part by the Department of
Energy under grants DE-FG06-90ER40561 and DE-FG03-97ER41014.


\end{document}